\documentclass[aps,prl,twocolumn,showpacs,superscriptaddress]{revtex4}
\usepackage{graphicx}
\usepackage{amssymb}

\newcommand{\beq}{\begin{equation}}
\newcommand{\eeq}{\end{equation}}
\begin{document}

\title{Correlated dynamics in human  printing behavior}

\author{Uli Harder}
\email{uh@doc.ic.ac.uk}
\affiliation{Department of Computing,  Imperial College London, London
  UK SW7 2AZ.} 
\author{Maya Paczuski} 
\email{maya@ic.ac.uk}
\affiliation{John-von-Neumann
Institute for Computing, Forschungszentrum J\"{u}lich, D-52425 J\"{u}lich,
Germany}
\affiliation{Department of Mathematics,  
Imperial College London, London UK SW7 2AZ.}

\begin{abstract}
  {Arrival times of requests to print in a student laboratory were
analyzed.  Inter-arrival times between subsequent requests follow a
universal scaling law relating time intervals and the size of the
request, indicating a scale invariant dynamics with respect to the size.
The cumulative distribution of file sizes is well-described by 
a modified power law often seen in non-equilibrium critical systems.
For each user, waiting 
times between their individual requests show long range dependence and
are broadly distributed from seconds to weeks. All results are incompatible
with Poisson models, and may provide
evidence of critical dynamics associated with
voluntary thought processes in the brain.
}

\end{abstract}

\pacs{89.75.-k, 89.75.Da, 02.50.E, 05.40.-a}

\maketitle

%%%%%%

Since the early work of Berger and Mandelbrot~\cite{berger} examining
error clustering in telephone circuits, it has been recognized that
standard Poisson models may be inadequate to describe electronic
information networks.  This was confirmed, for instance by Leland et
al.~\cite{LTWW}, who studied network traffic and found that packet
traces show scaling behavior.  
Observations of scaling behavior raise a number of
questions about how to model these systems, optimize performance, or
improve design. Significant effects include  an increase in response
times, required buffer sizes, etc. In Ref.~\cite{AESOP:mascots-2002-pe} 
the authors show how the file size distribution of a web server effects the
resulting network traffic.
Large fluctuations (which are inherent in critical systems)
 in packet traffic or
demand for resources in computer networks can significantly degrade
worst case performance~\cite{huberman}.
Scaling behavior has  been found not only in the size
distribution of files stored in computer systems~\cite{UFSD}, and the sizes
 of web server requests~\cite{C}, but also in the physical
structure of the internet~\cite{faloutsos} and the hyper-link
structure of the world-wide web
\cite{albert,bornholdt02:_book}.  So far, no definitive causes
have been established for the complexity of the modern information network.
Of course, humans interact when they build the internet, make hyper-link
connections, and send and receive information.  Like
traffic jams \cite{nagel_pac} on roads, internet jams are produced by
humans who act and react, often in response to
information originating within the network or outside it.  Various
parts of the information network/user system are themselves
complex systems, and one of the problems in modeling modern
information networks  is how to disentangle these effects.

One recognizes that psychological experiments have demonstrated that correlated
dynamics occurs in  individual human behavior
\cite{KMT93,CC94,DA96,gottschalk,paulus}, even in situations where
interactions with other humans are minimal~\cite{CDK,GTM,OHT03}. For
instance, Ref.~\cite{GTM} describes an experiment where
subjects had to estimate the duration of time intervals from
memory. The time series of errors in the estimates exhibits a $1/f$
power spectrum, showing that the errors are correlated in time. In
contrast, the sequence of reaction times to an event showed no long
range correlation. The authors proposed that  long range
dependence is associated with voluntary thought processes in the
brain~\cite{GTM}.  Similar observations were made for the dynamics of moods
\cite{gottschalk} and psychotic states.  For instance the
distribution of time intervals between subsequent hospitalizations for
schizophrenia is approximately power-law~\cite{DA96}. 
 A physical basis for these
behaviors may be related to scale-free functional networks in the brain, which
have recently been observed in situ~\cite{dante2}.

In order to better describe individual human behavior in a networked
computing environment, we study a simple case where the use
or demand is primarily associated with individual choice rather than with
group dynamics.  The particular quantity we focus on is the
inter-arrival times between subsequent print requests made by users in a computing
laboratory for university students.  We find evidence of long range
correlations in the inter-arrival times for individual users to send
requests, as well as a broad distribution of inter-arrival times. The totality
of print requests from all users reveals a scaling law relating
inter-arrival times and the sizes of the print request. This law indicates
that the same (re-scaled)
 dynamics is responsible for requests to print small and large
documents. This law is
 similar to that
recently observed for waiting times between successive
earthquakes~\cite{bak02:_unified,alvaro,davidsen_goltz} or solar
flares~\cite{bps}.  The
scaling function for the re-scaled inter-arrival times is approximately
log-normal.  The cumulative distribution of the sizes
of print requests is well-described by a modified power law, which is referred
to  as the $\chi^2$ distribution of
superstatistics~\cite{beck_cohen,beck}, or the q-exponential of non extensive
statistical mechanics
\cite{tsallis,wilk}.  An elementary stochastic
process is studied that reproduces some, but not all, of the observed
features. Our results are supportive of the hypothesis that the brain
operates at or near a self-organized critical state \cite{bak_brain}.
It also suggests the possibility of using data collected via  the modern information network
to systematically investigate models of human behavior.

The Department of Computing at Imperial College London maintains a
networked printing system for staff and students. The student labs
offer about 300 computer work spaces, and are divided into different
rooms, the largest one accommodating up to 150 students. The printers
are networked and accessible from any machine in the department.  A
user selects a printer and submits her print job to a central
server. The server records the time a request is submitted with a resolution
of one second.  It also records the size of the request,
the user name and the intended printer. This investigation focuses on
requests sent to the printer, {\em chrome}, that is located in the
largest room. The labs are closed between 23:00 and 7:00, but users can
print during closure times when logged in remotely.  The data used here include the entire
year of 2003 and closure
times have been included in the analysis.  Table 1 gives relevant
parameters for the data set studied, which can be accessed
at~\cite{data}.
\begin{table}[htbp]
  \centering
  \begin{tabular}{|l|l|}
   \hline
    number of users & 1122\\ \hline
    number of users issuing & \\ 
    more than three requests & 1001\\ \hline
    number of requests per year & 73853\\ \hline
    mean document size & 1.2 Mbytes\\
    mean time between requests & 7.1 min\\ \hline
    minimum time resolution & 1.0 sec\\ \hline
  \end{tabular}
  \caption{Parameters of the user and printing system in 2003.}
  \label{tab:key}
\end{table}

%525600

%\section{Analysis}

\begin{figure}[htb]
    \centering
     \includegraphics[width=6cm,angle=270]{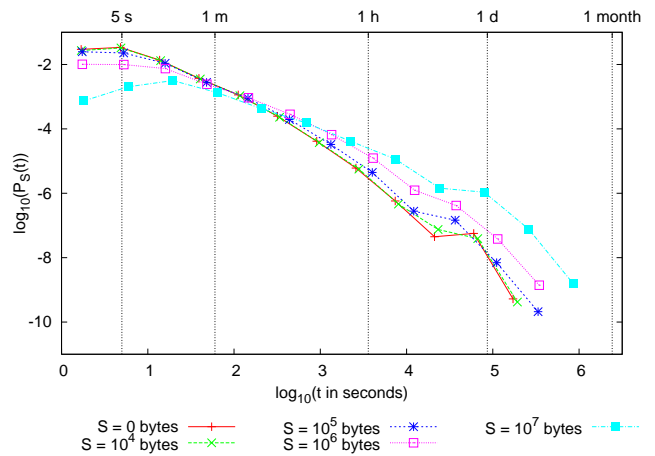}
     \caption{Distribution of inter-arrival between subsequent requests to
       the printer "chrome" in 2003.  Different curves are for
       different threshold sizes of the requests. }
     \label{fig:unscaled-iat}
 \end{figure}

We first analyze the distribution of  inter-arrival times
 between subsequent print
requests for the entire year. Time
differences from the logged event times $T^S_i$ are measured as
\begin{equation}
  \label{eq:diff}
  t_i^S  = T^S_{i+1} - T^S_i, \text{where } 0 \leq i
  \leq N^S \quad .
\end{equation}
The superscript $S$ refers to the size of the print request in bytes
and indicates that this set of times only includes requests that
are larger than $S$. The quantity $N^S$ is the number of print requests that are
larger
than $S$. Time intervals of length zero are
neglected from the analysis. For each chosen threshold $S$ we estimate
 $P_S(t)$, which is the probability
of a certain time interval $t$ between subsequent requests
of size $S$ or larger.  To display this distribution we count the
number of time differences in exponentially growing bins and normalize
the count by the bin size. Fig. \ref{fig:unscaled-iat} shows that the 
shape of the waiting time distribution depends on the size threshold,
$S$, of the documents. This could indicate  different dynamical processes
responsible for the small and large documents.  However, all distributions are  broad and show an
anomaly near one day. The anomaly is related to the overnight
closure of the  labs.

\begin{figure}[htb]
    \centering
     \includegraphics[width=6cm,angle=270]{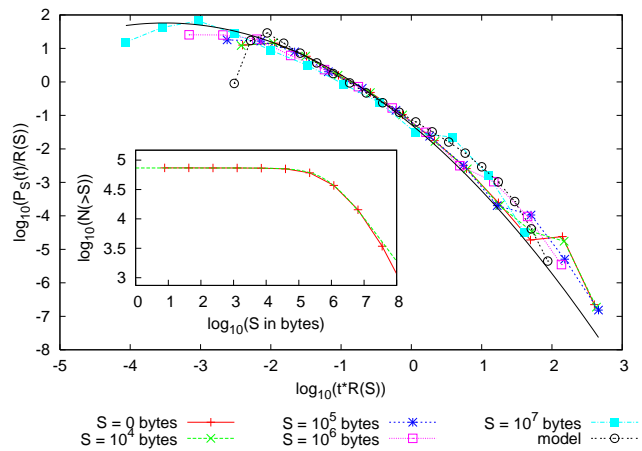}
     \caption{Universal scaling law for the inter-arrival
times between requests larger than size $S$, according to Eq. \ref{eq:ansatz}.
The solid line is a fit of to a log-normal function as described in the
text. Data from the numerical simulation is also shown. The inset displays the cumulative distribution of requests sizes.}
     \label{fig:scaled-iat}
 \end{figure}

To determine if a different dynamics is responsible for requests of different
sizes, we implement
 a scaling argument similar to one recently put forward by Bak {\it et
   al}~\cite{bak02:_unified} to describe the waiting time statistics
   of earthquakes.     The
   average time between requests $\langle t \rangle_S$ may provide a 
    rescaling factor for the inter-arrival times, so that
   the distributions measured with different
   size thresholds, $S$, collapse onto a single scaling function.
   Of course, $\langle t\rangle
   _S = \frac{T}{N_{>S}}= \frac{1}{R(S)}$.  Here $T$ is the time span
   of the record and $R(S)$ is the rate of requests larger than $S$.  $N(>S)$ is the cumulative number of requests
   larger than size $S$. As
   shown in the inset of Fig.  \ref{fig:scaled-iat},  $N(>S)$ is
well described by a modified power law~\cite{beck_cohen,beck,tsallis,wilk}:
\begin{equation}
  \label{eq:S}
  N(>S) \sim \frac{1}{(1+(S/S^*))^{\gamma -1}}
\end{equation}
  where
    $S^* = (7.9 \pm 0.5) \times 10^{5}$ and
   $\gamma-1=0.76 \pm 0.03$.  

 We test
  the {\it ansatz} 
\begin{equation}
  \label{eq:ansatz}
 P_S(t)\sim R(S)g(t R(S)) \quad ,  
\end{equation}
 where $g(x)$ is a scaling function and $x=tR(S)$ is a scaling variable.
Fig.~\ref{fig:scaled-iat} shows the results of rescaling the different
 curves in Fig.~\ref{fig:unscaled-iat} by their average rate.  We see that the scaling
{\it ansatz} of Eq. \ref{eq:ansatz} appears to hold over a wide range, about seven orders of
magnitude in the scaling variable. This indicates that the same scale invariant
dynamics operates when users send requests of any size.
 The slight
 deviation from data collapse at short times is due to the finite
 temporal resolution of our data (one second). There is an additional
 deviation due to the  diurnal period. 
The scaling function $g$ is close to  a log-normal distribution: 
\begin{equation}
  \label{eq:lognormal}
  g(x) = \frac{ 1 }{ \sqrt{2 \pi} \sigma x   }  
\exp(-\frac{ (\ln(x) - m)^2}{2 \sigma^2 } ) 
\end{equation}
with $m= -3.41 \pm 0.07$ and $\sigma = 2.16 \pm 0.04$, as also shown
 in Fig.~\ref{fig:scaled-iat}. This feature is also found in numerical simulations
of a stochastic process described later.

\begin{figure}[htb]
    \centering
     \includegraphics[width=6cm,angle=270]{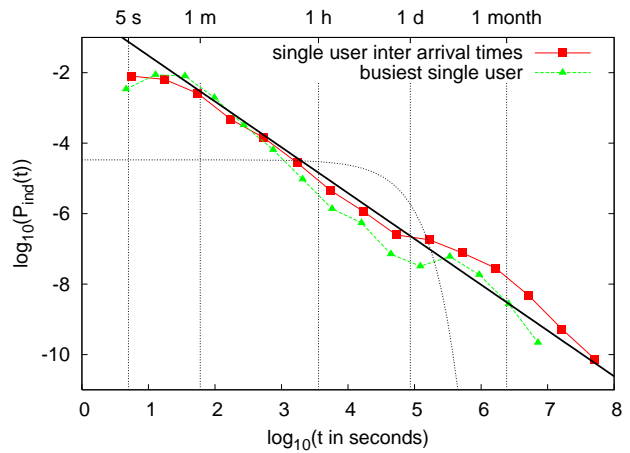}
     \caption{Single user inter-arrival time distribution, averaged over all
users and for the single busiest user. The solid, straight line indicates a power law
distribution, $P_{\rm ind}(t) \sim t^{-\alpha}$ with $\alpha=1.3$. For comparison,
an exponential distribution with the same rate as the busiest user is 
shown as a dashed curve.}
     \label{fig:single-iat}
 \end{figure}

The inter-arrival times for all users do not
 necessarily give a good estimate for the times that pass between subsequent
requests issued by a {\em single} user. To this end we study
the inter-arrival times $t_i^u$ for each user $u$ printing more than three
documents over the one year period. In the discussion below we set the threshold
$S=0$.
\begin{equation}
  \label{eq:user}
  t_i^u  = T^u_{i+1} - T^u_i, \text{where } 0 \leq i
  \leq N^u .
\end{equation}
Each user's list of inter-arrival times is concatenated to determine
the probability  $P_{\rm ind}(t)$ of single user inter-arrival times,
shown in Fig.~\ref{fig:single-iat}.  This distribution is
approximately a power law over several decades ranging from one minute
to about a day, with an exponent $\alpha
\approx 1.3$. We also analyze the inter-arrival times for the
busiest single user, which is similar.  For comparison we
show in Fig.~\ref{fig:single-iat} an exponential distribution for a
Poisson event process that has the same average rate,
$\lambda=3.4\times 10^{-5}/{\rm sec}$, as the process of the busiest
single user.  A critical system with a  power-law distribution of intervals
 is a more
accurate description of the data than a Poisson model of print requests.

To decide if inter-arrival times  are correlated, we measured
the auto correlation function of waiting times
for  single users.
The autocorrelation $a_u(\tau)$ at lag step $\tau$ is defined as
\begin{equation}
  \label{eq:auto}
  a_u(\tau) = \frac{1}{N_u - \tau} \sum_{i=1}^{N_u - \tau} s^u_i s^u_{i+\tau}
\end{equation}
where $s^u_i = t_i^u - \frac{1}{N_u} \sum_{j=1}^{N_u} t_j^u$.  If the
inter-arrival times are uncorrelated and independent, the arrival process of individual
requests to print can be modeled as a fractal renewal process
\cite{pmb,LT193}.  Analyzing data separately for the three most busy users, we
find that the auto correlation function decays as $1/\tau^{\delta}$ with
$\delta \approx 0.6$. When the order of the inter-arrival times
for an individual user are shuffled randomly this power law
disappears, and the waiting times become uncorrelated, with $a_u(\tau)$ independent
of $\tau$ for $\tau \geq 1$.   The sequence of inter-arrival times
 for individual users are
correlated over the entire time span of our data set.

Our data shows that models of criticality are relevant for describing individual
human
behavior in the modern information network.  Lacking, at present, a microscopic
dynamical model, we compare our observations with results from a 
simple stochastic process. Consider $N$ arrival streams of print
requests. In each stream,  time intervals between subsequent
requests are independent random variables chosen from a truncated
Pareto distribution.  We neglect correlations between
intervals.  All intervals have
the same probability distribution
\begin{equation}
  \label{eq:pareto}
  P_{\rm ind}(x) = \frac{1}{C} k x^{-1-k} \text{ where } 1 \leq a \leq x \leq b
\end{equation}
where $a$ and $b$ are the points where the Pareto distribution is
truncated and $C$ is a normalization constant.  We choose the
parameter $k=0.3$ motivated by the  results in Fig.
\ref{fig:single-iat}. The short time cut-off $a=2.5$ sec is set to
reflect the fact that in some application users must wait
 before a subsequent print job can be sent off.  Most students
 leave after  at most 8 years, so $b = 8$ years appears to
be a reasonable choice. Generating approximately 73,000 requests in
a year fixes the number of users close to $N = 1000$.
 
At the start of the numerical
 simulation we schedule an arrival event for each stream
according to  Eq. \ref{eq:pareto}. Upon each arrival, the next arrival time
is scheduled using the same distribution. The system
takes  about 5 years with the above parameters to reach a statistically
stationary state.
 As shown in Fig. 2, the inter arrival times measured in the simulation
compare fairly well with the real data. 
However the real data has significantly larger variance.
 
 We also examined the  time series defined by the number of print requests
 in each second.  We calculated the power spectrum $S(f)$ of this time series
 and find $1/f^{\alpha}$ behavior,
 as shown in Fig. 4.
The exponent $\alpha$ observed in the numerical simulation is fixed by the
value of $k$
 in Eq. \ref{eq:pareto}, and is $\alpha =
 0.3$ \cite{PMB96,LT193}. The real data 
show instead a larger value $\alpha \approx 0.5$,
  which indicates, just as
the auto-correlation function $a_u(\tau)$, that the real
arrival process is more complicated than a fractal renewal process. A more
accurate model of individual user behavior in a computing network may be that of Davidsen and Schuster~\cite{joern1}.
\begin{figure}[htb]
    \centering
     \includegraphics[width=6cm,angle=270]{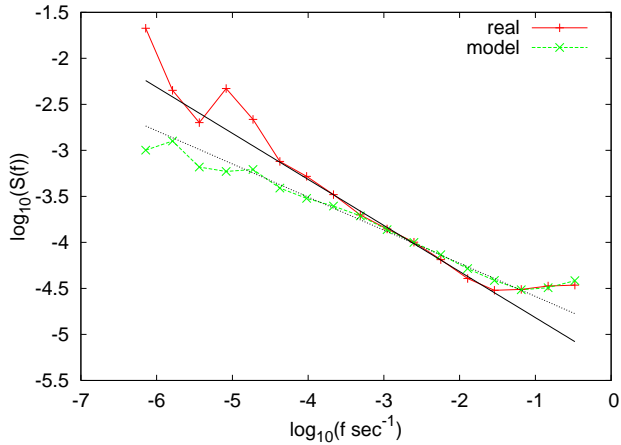}
     \caption{Power spectrum of the time series defined by the print
     requests per second  
     based on the 
     real arrival data and the 
     simulated arrivals in the fifth 
     year. The solid line is a fit for the real data,
     the dashed one for the simulation results, see text.} 
     \label{fig:sim-power}
 \end{figure}

The authors  thank the Computer Support Group of the
Department of Computing at Imperial College, especially T.
Southerwood, for making the data available to us.
M.P. thanks J. Davidsen, A.L. Stella, and P. Grassberger for  conversations.
U.H. would like to thank A. Argent-Katwala, T. Field and W.
Knottenbelt for interesting discussions and suggestions. U.H. is
funded by  EPSRC (research grant PASTRAMI, GR/S24961/01)
%\bibliographystyle{apsrev}
%\bibliography{queue_prl}

\end{document}